\documentclass{article}

\PassOptionsToPackage{numbers, compress}{natbib}

\usepackage[preprint]{neurips_2025}

\usepackage[utf8]{inputenc} 
\usepackage[T1]{fontenc}    
\usepackage{hyperref}       
\usepackage{url}            
\usepackage{booktabs}       
\usepackage{amsfonts}       
\usepackage{nicefrac}       
\usepackage{microtype}      
\usepackage{xcolor}         

\usepackage{graphicx}
\usepackage{subcaption}
\usepackage{wrapfig}
\usepackage{listings}
\usepackage{caption}             
\usepackage{multirow}

\title{NPUEval: Optimizing NPU Kernels with LLMs and Open Source Compilers}

\author{%
  Sarunas Kalade\\
  Advanced Micro Devices\\
  \texttt{sarunas.kalade@amd.com} \\
  \And
  Graham Schelle \\
  Advanced Micro Devices \\
  \texttt{graham.schelle@amd.com} \\
}

\begin{document}

\maketitle

\begin{abstract}


Neural processing units (NPUs) are gaining prominence in power-sensitive devices like client devices, with AI PCs being defined by their inclusion of these specialized processors. Running AI workloads efficiently on these devices requires libraries of optimized kernels. Creating efficient kernels demands expertise in domain-specific C++ with vector intrinsics and in-depth knowledge of the target architecture. Unlike GPU programming, which has had years to mature, NPU programming is new, with smaller and more fragmented developer communities across hardware platforms. This fragmentation poses a challenge when utilizing LLMs to assist in writing NPU kernels, as domain-specific optimized code examples are underrepresented in LLM pre-training data.

In this paper we introduce NPUEval -- a benchmark for writing and evaluating NPU kernels, consisting of 102 common operators for machine learning workloads. We evaluate LLM generated code on actual hardware based on both functional correctness and vectorization efficiency using open source compiler tools targeting the AMD NPU. We evaluate a range of state-of-the-art LLMs with a mix of proprietary and open-weight models. Latest reasoning models like DeepSeek R1, show promising results achieving out-of-the-box 50\%+ vectorization on select kernels. However, the average score across the entire dataset remains roughly 10\% even with compiler feedback and vectorized kernel examples -- showing that this is a challenging dataset even for frontier models. The dataset and evaluation code will be released with a permissive open source license, providing an essential benchmark for advancing research in code generation and NPU kernel optimization.
\end{abstract}

\section{Introduction}

Large language models (LLMs) have become compelling code generation assistants, achieving breakthrough performance on various coding benchmarks \cite{humaneval}\cite{mbpp}\cite{evalplus}\cite{cassano2023multipl}\cite{chai2024mcevalmassivelymultilingualcode}\cite{gong2024evaluation}\cite{li2024evocodebenchevolvingcodegeneration}\cite{zhuo2024bigcodebench}. More challenging benchmarks that focus on issue solving like SWE-Bench\cite{jimenez2023swe} are now also being tackled at over 62\% success rate by frontier models, and rapidly improving\cite{sonnet_3_7_swebench}. However, while these models excel on typical Python tasks, many benchmarks prioritize pass/fail metrics over code quality. A poorly written algorithm can still pass functional tests, but may not be suitable for production systems. This is particularly problematic when generating efficient kernels for hardware accelerators -- unoptimized kernels are not useful in accelerated applications.

There has been a surge of NPUs to power AI acceleration workloads from various silicon providers\cite{amd_xdna}\cite{apple_neural_engine}\cite{huawei_ascend}\cite{intel_npu}\cite{samsung_npu}\cite{qualcomm_hexagon}. However, related work on optimized kernel generation have been primarily targeting GPUs\cite{kernelbench}\cite{lange2025ai}\cite{li2025tritonbenchbenchmarkinglargelanguage}. GPU programming languages and ecosystem have had years to mature and are well represented in LLM pre-training data. For NPU programming, on the other hand, the developer communities are smaller and more fragmented across hardware architectures, with less mature software stacks. We found that LLMs often struggle to produce optimized solutions for vendor specific code, even though frontier models are good at solving general software problems.

This paper introduces NPUEval, a dataset designed to evaluate LLMs' ability to generate vectorized kernel code for AMD NPUs. The accessibility in client devices makes these accelerators an attractive platform for kernel code generation research, maximizing reproducibility by running on commodity laptops and miniPCs. The entire evaluation harness, including the compiler, is based on open source tools and we are releasing the dataset and associated code under a permissive license. The dataset is composed of prompts, behavioral models, and data movement information for each kernel. Our evaluation framework measures LLM performance on functional correctness (similar to benchmarks like HumanEval\cite{humaneval}) and cycle-accurate performance metrics such as vectorization score (determined by the percentage of cycles spent executing vector instructions).

Our results show that while LLMs can generate functionally correct code that run on NPU hardware, and with the addition of simple techniques like compiler feedback and RAG they are capable of writing vectorized implementatons.

The contributions of this paper are:
\begin{itemize}
    \item Introduction of NPUEval, which to our knowledge is \textbf{the first benchmark for evaluating LLMs in generating vectorized code for NPU kernels}.
    \item A \textbf{fully open source stack} (compiler and driver) for programming AMD NPUs focused on single kernel development that can run on laptops.
    \item A comprehensive evaluation harness providing correctness evaluation, \textbf{cycle-accurate performance metrics}, and kernel microcode outputs that can be built upon using agentic workflows for further optimization.
    \item A \textbf{reference LLM pipeline} that utilizes compiler feedback and retrieval augmented generation (RAG) of vectorized kernel examples to steer the LLM outputs towards vectorized solutions and reduce hallucinations.
\end{itemize}

\section{Vectorization}

The hardware targeted in this paper is the NPU found on latest AMD laptops and miniPCs like the Phoenix, Hawk Point or Strix Point based machines, which internally use AIEs to perform kernel computations. A device like Phoenix will have 20 AIE tiles which are individual compute units capable of independently processing data. Each AIE tile has a vector processing unit (VPU) as well as a scalar unit, which makes these compute units quite flexible since the scalar unit can execute arbitrary C++ code.

These AIE tiles require kernels just like a GPU, and writing them using optimized, vectorized code is essential to fully leverage the hardware architecture for maximum performance. Inefficient kernels will dramatically increase latency and power consumption, which completely undermines the purpose of having AIEs included in these devices. However, in order to squeeze out as much performance as possible we want to maximize the computation on the VPU, which means writing vectorized code by utilizing C++ AIE APIs and intrinsics.

\lstset{
  basicstyle=\ttfamily\small,
  language=C++,
  keywordstyle=\color{blue},
  commentstyle=\color{gray},
  stringstyle=\color{orange},
  breaklines=false,           
  tabsize=2,
  columns=flexible,
  keepspaces=true,
  showstringspaces=false,
  xleftmargin=0pt,
  linewidth=\linewidth,
  aboveskip=0pt,
  belowskip=0pt
}

\begin{figure}[!h]
\centering

\newlength{\codefigheight}
\setlength{\codefigheight}{4.6cm} 

\begin{minipage}[t][\codefigheight][t]{0.48\textwidth}
\begin{lstlisting}
void passthrough(uint8_t *in_buffer, 
                 uint8_t *out_buffer){
    uint32_t nbytes = 512;
    for (int i=0; i<nbytes; i++) {
        out_buffer[i] = in_buffer[i];
    }
}
\end{lstlisting}
\vfill
\subcaption{Non-vectorized (scalar) code}
\label{fig:scalar}
\end{minipage}
\hfill
\begin{minipage}[t][\codefigheight][t]{0.48\textwidth}
\begin{lstlisting}
void passthrough(uint8_t *in_buffer, 
                 uint8_t *out_buffer){
    uint32_t loop_count = 8; // 512/64

    for (int i = 0; i < loop_count; i++) {
        buffer = ::aie::load_v64<>(in_buffer);
        ::aie::store_v(out_buffer, buffer);
        in_buffer += 64;
        out_buffer += 64;
    }
}
\end{lstlisting}
\vfill
\subcaption{Vectorized 64-byte implementation}
\label{fig:vector}
\end{minipage}

\caption{Simple example of vectorizing a passthrough kernel}
\label{fig:code-comparison}
\end{figure}

A very simple example showcasing vectorization is in Fig. \ref{fig:code-comparison}. The scalar code (Fig. \ref{fig:scalar}) is a passthrough kernel that naively iterates one element at a time in a for loop and copies the input to the output buffer. The vectorized code in Fig. \ref{fig:vector} is the same kernel, but written to utilize the VPU using vector instructions - this will process a chunk of data at a time rather than single elements, significantly improving throughput.

Optimized kernels for vectorization can look more complicated and result in more lines of code than this simple example, but the vectorization methodology remains the same (see Appendix \ref{appendix_a} for reference vectorized code). The challenge for the LLMs will be to generate vectorized code that is both correct and efficient in using the VPU by utilizing the correct APIs.

\section{The Dataset}

\begin{figure}[!h]
    \centering
    \includegraphics[width=.9\textwidth]{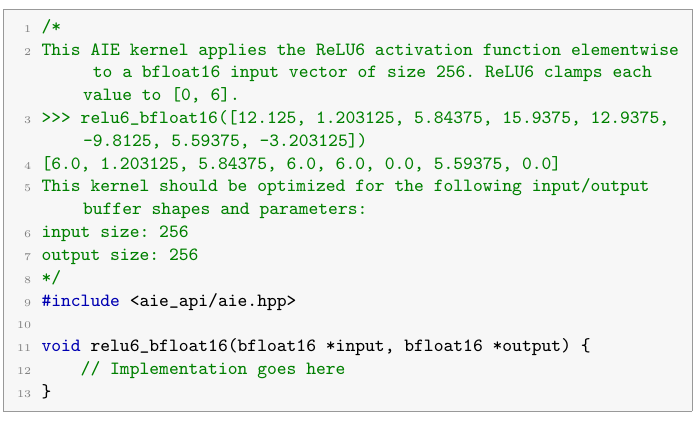}
    \caption{Sample prompt from the NPUEval dataset.}
    \label{fig:prompt_sample}
\end{figure}

The NPUEval dataset is a collection of prompts consisting of AIE kernel definitions accompanied with docstrings containing the kernel description, input/output examples, and anticipated data movement and runtime parameters (see Appendix \ref{appendix_b} for more detail). An example prompt is shown in Fig. \ref{fig:prompt_sample}.

\subsection{Structure}

Each kernel in the dataset will have the following components:
\begin{itemize}
    \item Prompt - HumanEval inspired prompts containing a basic explanation of what the kernel does and function signature.
    \item Data movement information - specifying the data sizes coming in and out of the tile.
    \item Behavioral model - NumPy-based Python implementation of the target kernel.
    \item (Optional) Canonical C++ solution. While the Python solution is used for correctness evaluation, the canonical solution is useful for regression testing of the harness.
\end{itemize}

To generate optimized kernels that fully utilize the target hardware, kernels may be written for specific fixed input data sizes. For this reason we include data movement information for each kernel, specifying the data sizes coming in and out of the tile.\cite{hunhoff2025efficiencyexpressivityextensibilityclosetometal}

\subsection{Dataset considerations}
Since the main task is writing code for specialized hardware there are some considerations and challenges when building this dataset that are not typically considered in LLM coding benchmarks.

\subsubsection{Data type support}
In addition to common NPU data types like int8 typically used to accelerate quantized neural networks, AMD's NPU supports bfloat16 which is relatively new. Python libraries like NumPy don't have native support for these newer ML data types yet. To generate bfloat16 behavioral models we used Google Jax's ml\_dtypes\cite{ml_dtypes} library which provides a NumPy-compatible bfloat16 type. The NPU has a programmable rounding mode for its bfloat16 implementation, which we leverage to ensure the same operations will result in identical outputs minimizing compounding rounding errors.

\subsubsection{Floating point kernel evaluation}
Floating point kernel solutions may not match the Python behavioral model outputs exactly. The same problem can be solved using different algorithms with slight precision tradeoffs. For example, when estimating trigonometric functions like \texttt{sin}, the NPU might use a different polynomial approximation than the CPU implementation used in the behavioral model. Floating point operations are also not associative, so doing operations in a different order will produce slightly different values. When evaluating correctness of the generated kernels we use a large default absolute error threshold of 1e-2 (similar default to related kernel generation works\cite{kernelbench}). This default is fine for logical and integer operators, but for lower precision floating point operations the tolerance is set to 2e-2 to give the LLMs some leeway in the algorithmic search space. More complex kernels like tanh or softmax activations have their tolerances set to 3e-2.

\subsubsection{Data movement}
We have pre-defined data movement for each kernel in order to compile the AIE graph using MLIR-AIE tools\cite{mlir_aie}, this is included in the prompts. The test vectors are used to infer the data movement and automatically generate underlying MLIR code to configure the device. In dataflow programming kernels are aware of data sizes coming in and out of the tile\cite{laan2024developingblaslibraryamd}. This is essential for the compiler to be able to optimize buffer allocation and bank assignment.\cite{hunhoff2025efficiencyexpressivityextensibilityclosetometal}

\subsubsection{Hardware availability}
Most code generation benchmarks that set out to solve problems in Python or other high-level languages are convenient because they can run on any commodity hardware. Writing code for hardware accelerators is difficult because you need to be able to run tests on-target or use a simulator (typically much slower, limiting the number of iterations). While AIEs can also be found on development boards for embedded applications\cite{AMD_VCK190}, these platforms are less accessible to mainstream users. We target AMD NPUs as they are readily available on consumer hardware, making reproduction of our results more approachable to a wider range of researchers. The results presented in this work were obtained using a laptop with a Ryzen 9 7940HS chip.

\section{Evaluation}

The generated kernels are evaluated using the following criteria:

\begin{itemize}
    \item \textbf{Compilation} - does the kernel compile? Is it syntactically correct C++ code using valid NPU vector unit API calls and intrinsics?
    \item \textbf{Functional correctness} - does the kernel produce the correct output for the given input (given a provided error tolerance)?
    \item \textbf{Performance} - how long does the kernel take to execute and how efficiently is the VPU utilized?
\end{itemize}

\begin{figure*}[!h]
    \centering
    \includegraphics[width=\textwidth]{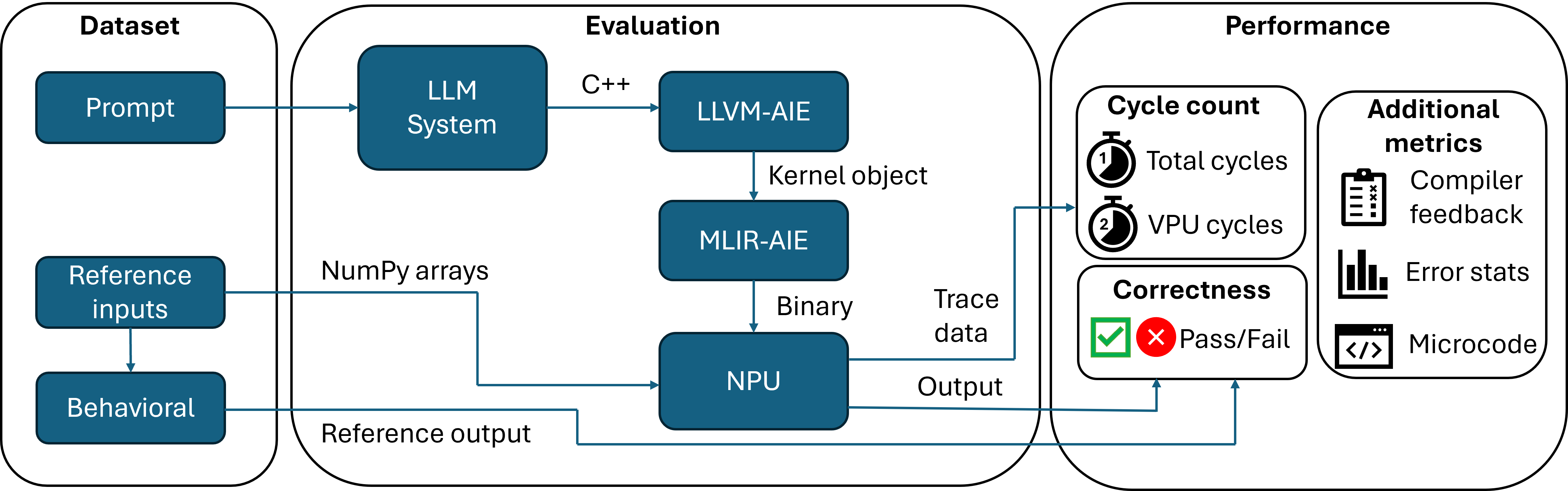}
    \caption{Overview of NPUEval evaluation pipeline.}
    \label{fig:eval_overview}
\end{figure*}

\subsection{Evaluation harness setup}

A high-level overview of the evaluation pipeline is shown in Fig. \ref{fig:eval_overview}. The dataset includes kernel prompts, behavioral models and test vectors. The evaluation harness compiles the generated C++ kernel code and runs it on-target. The outputs are then compared against the expected simulation outputs from the Python behavioral models.

\textbf{Compiler}.To evaluate the LLM generated code we use the LLVM-AIE compiler\cite{llvm_aie}, which is a fork of LLVM specifically used for AIE kernel programming. This tool is entirely open source and can be installed from its GitHub repository. To run agents that are generating compilable code you only need an x86 machine to reproduce the code generation steps with compiler feedback - evaluation whether the code is syntactically correct is still useful even if one does not have access to a machine that has an NPU.

\textbf{Application builder}. When programming the NPU, the AIE array needs to be configured and the kernel loaded onto the right AIE tile. We utilize the open source MLIR-AIE\cite{mlir_aie} framework and IRON\cite{hunhoff2025efficiencyexpressivityextensibilityclosetometal} bindings for this task. The NPUEval evaluation harness has templated graphs for most kernel data types and input/output port numbers. The data movement passed onto MLIR-AIE is determined by parsing the test vectors and behavioral model outputs, which are part of the dataset.

\textbf{Runtime}. To move data in and out of the NPU and execute the kernels, we use Python bindings to the NPU driver. This integration allows us to easily work with NumPy arrays, and the behavioral model can be quickly checked against NPU outputs.

\textbf{Performance metrics}. In addition to the core evaluations, we collect supplementary information about the generated kernels. This includes metrics like execution time, and accuracy measurements. For accuracy, we compute both maximum absolute error (the numerical difference between each output value and reference value) and maximum relative error (the percentage difference relative to the reference value).

Agents can utilize this information to iterate on the code generation process. For example, if the kernel is failing the functional test, seeing the maximum absolute error compared to the reference Python output can help refine the algorithm. If the kernel is taking too long to run and the VPU is not being utilized, the agent could try different vectorization strategies.

\textbf{Post-processing}. This is essential in LLM evaluation since the responses from some LLMs can sometimes not be immediately usable for automatic evaluation. These models will often mix in english language to the responses and denote codeblocks using markdown markers like "```". LLMs also have a tendency to output a main() function to test the kernel. We use regex to extract codeblocks embedded in markdown and truncate any extraneous functions.

\section{Generation}


Along with the dataset and evaluation harness, we provide a reference code generation pipeline that can be used with a variety of LLMs. A system prompt is provided to guide the LLMs into generating self-contained C++ solutions, we generate a vector database of open source AIE kernels for RAG and provide compiler feedback to the LLM to reduce errors and hallucinations.

\subsection{System Prompts}
A system prompt will typically be used to guide the model's behavior by providing context and expected outcomes. Based on initial experimentation, adding a system prompt was imperative to get well formatted code from many of the LLMs. Some models will attempt to break down the solutions into multiple codeblocks, explanations and produce overly verbose outputs, which make it difficult to automatically parse the solution. The prompt used for NPUEval is shown below.

\definecolor{lightgray}{rgb}{0.95, 0.95, 0.95}

\lstset{
  basicstyle=\tiny\ttfamily,
  backgroundcolor=\color{lightgray},
  breaklines=true,
  frame=single,
  numbers=left,             
  numberstyle=\tiny\color{gray},  
  numbersep=5pt,            
  xleftmargin=15pt,         
  framexleftmargin=15pt,    
  captionpos=b,             
  stepnumber=1,             
  showstringspaces=false,    
  lineskip=2pt
}

\vspace{5pt}
\begin{lstlisting}[label=ref:system_prompt]
You are a part of a code generation system for AIE (AI Engines).

* Your job is to write C++ code for a single kernel that will run on an AIE tile.
* Produce only the C++ code for the requested kernel including any required headers and imports.
* Make sure the C++ code is complete and self contained in a single code block.
* Name the function exactly as specified in the request, and output only the kernel (no main(), examples, explanations or extra code).
\end{lstlisting}

While the system prompt does not guarantee that the model will follow the rules and generate self-contained C++ kernel code blocks, empirically we found it very helpful. Some models are not as good at following the directions, in which case more post-processing is still required (e.g., for unwanted main() function pruning).

\subsection{RAG}
The RAG database is composed of kernels drawn from open source GitHub repositories\cite{mlir_aie}\cite{riallto}. The kernels have been manually modified to exclude any scalar implementations, leaving only vectorized code examples.

We use llama\_index \cite{llama_index} to manage the vector database and retrieval of examples. The embeddings model is OpenAI's text-embedding-ada-002 (llama\_index default). For each test prompt two examples are provided to the LLM. RAG systems are a rich optimization space in itself\cite{gao2024retrievalaugmentedgenerationlargelanguage}, however it is out of scope for this paper -- we leave this for future work, and use llama\_index defaults for the evaluations.

\subsection{Compiler feedback}
The same compiler, LLVM-AIE, is used for generation as for evaluation. The generated kernel code is passed through the compiler and if it fails to compile, the error message is fed back to the model. Each LLM is allowed to retry code generation up to ten times. Results could potentially be improved even further with more compilation attempts, however we start seeing diminishing returns after ten - please see Appendix \ref{appendix_c}.

\subsection{LLM settings}
Where available we try to lock down random seeds and make the results as reproducible as possible. For OpenAI models we set the random seed to 42. For all models we set temperature to 0.0 and top\_p to 1.0. For reasoning models the temperature will be set to the default 1.0 and system prompt passed as part of the user message.

\section{Results}

Popular frontier LLMs were evaluated on the dataset, including latest versions of OpenAI's GPT-4.1 and Anthropic's Claude 3.7 Sonnet. We also include results of open-weight models like Meta's Llama family. And other open source models from DeepSeek that are known to well work for code generation tasks like DeepSeek R1.

\subsection{Out-of-the-box LLM evaluation}

\begin{figure}[!h]
    \begin{subfigure}[b]{.5\textwidth}
        \centering
        \includegraphics[width=\textwidth]{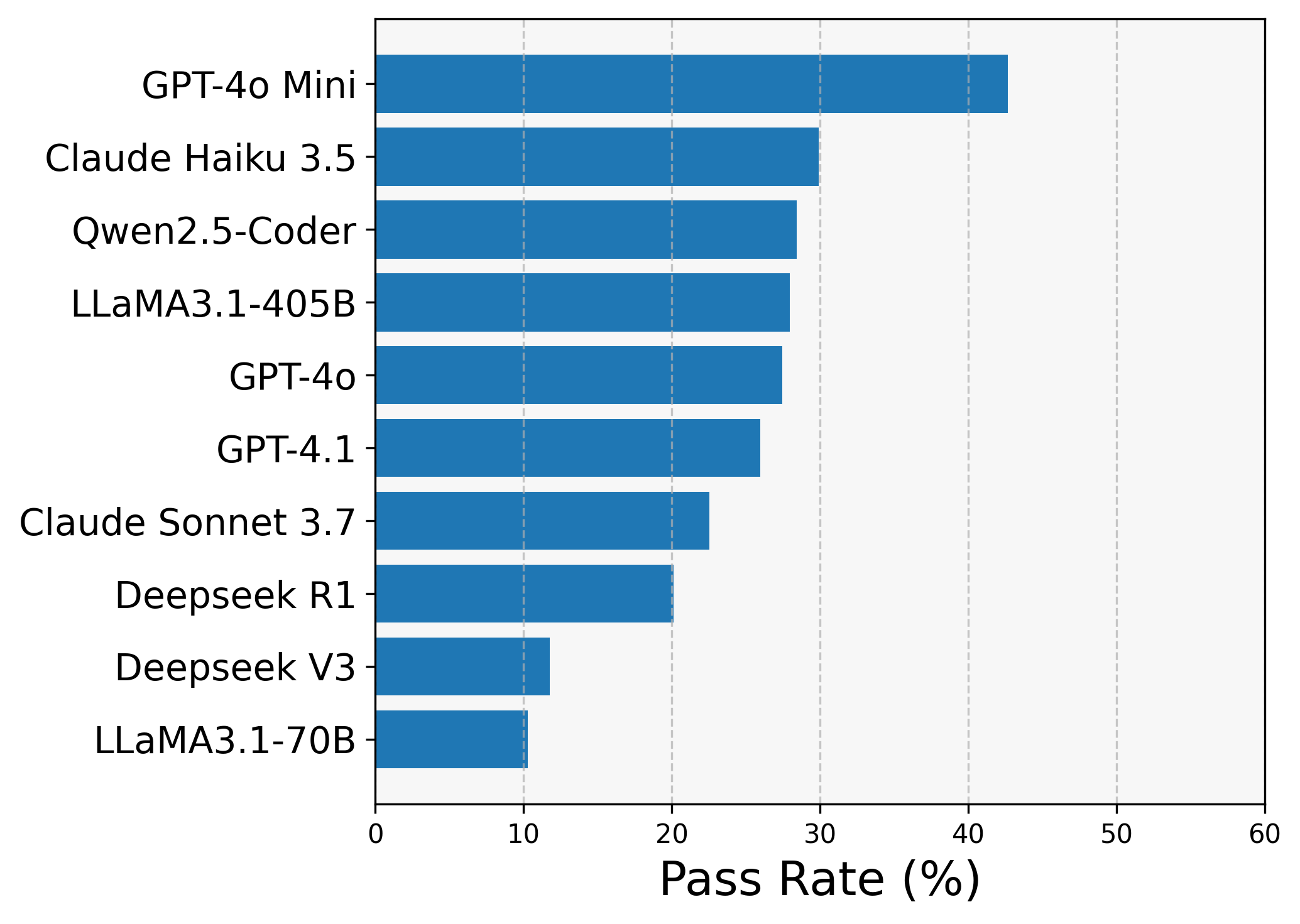}
        \caption{Functional correctness}
        \label{fig:outofbox_functional}
    \end{subfigure}
    \hfill
    \begin{subfigure}[b]{.5\textwidth}
        \centering
        \includegraphics[width=\textwidth]{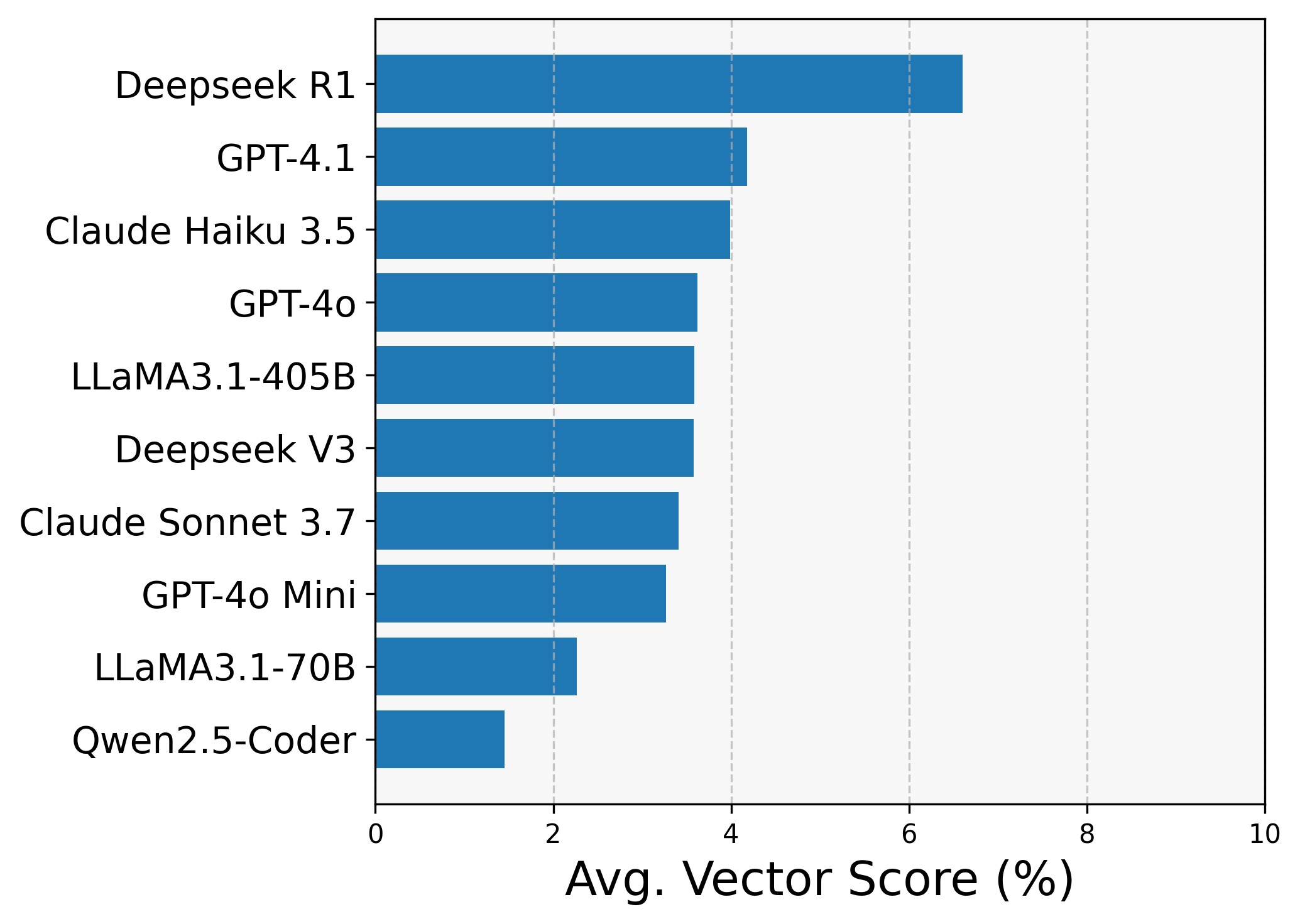}
        \caption{Vectorization score}
        \label{fig:outofbox_vector}
    \end{subfigure}
    \caption{Zero shot results (ranked from top to bottom)}
    \label{fig:outofbox_results}
\end{figure}

We first set the baseline for the models by testing them on the prompts directly only adding the system prompt outlined in \ref{ref:system_prompt}. The baseline results are summarized in Figure \ref{fig:outofbox_results}

\subsection{Correctness}

As shown in Fig. \ref{fig:eval_overview}, correctness is evaluated by comparing the kernel output of the programmed NPU with the reference output produced by the Python behavioral model. The effectiveness of compiler feedback is explored with a maximum of five recompilations. The summary of our results is shown in Table. \ref{tab:combined_results}.

\begin{table}[t]
\small
\setlength{\tabcolsep}{4pt}
\centering
\caption{Functional test pass rates (\%) with iterative re-compilation.}
\label{tab:combined_results}
\begin{tabular}{l|rrrr|rrrr}
\toprule
\multirow{2}{*}{Model} & \multicolumn{4}{c|}{No RAG} & \multicolumn{4}{c}{RAG} \\
\cmidrule(lr){2-5} \cmidrule(lr){6-9}
Num. recompilations & 0 & 1 & 2 & 5 & 0 & 1 & 2 & 5 \\
\midrule
Claude Haiku 3.5  & 34.3 & 44.1 & 52.9 & 61.8 & 25.5 & 40.2 & 49.0 & 53.9 \\
Claude Sonnet 3.7 & 21.6 & 52.9 & 64.7 & \textbf{73.5} & 23.5 & \textbf{47.1} & \textbf{59.8} & \textbf{70.6} \\
Deepseek R1       & 20.6 & 30.4 & 34.3 & 38.2 & 19.6 & 23.5 & 26.5 & 29.4 \\
Deepseek V3       & 0.0 & 42.2 & 55.9 & 60.8 & 23.5 & 27.5 & 31.4 & 39.2 \\
GPT-4.1           & 29.4 & 52.9 & 63.7 & 71.6 & 22.5 & 37.3 & 45.1 & 58.8 \\
GPT-4o            & 36.3 & 40.2 & 43.1 & 49.0 & 18.6 & 28.4 & 34.3 & 41.2 \\
GPT-4o Mini       & \textbf{58.8} & \textbf{65.7} & \textbf{66.7} & 66.7 & \textbf{26.5} & 34.3 & 34.3 & 35.3 \\
LLaMA3.1-405B     & 38.2 & 46.1 & 50.0 & 56.9 & 17.6 & 21.6 & 26.5 & 30.4 \\
LLaMA3.1-70B      & 11.8 & 37.3 & 42.2 & 52.0 & 8.8 & 17.6 & 22.5 & 23.5 \\
Qwen2.5-Coder     & 50.0 & 56.9 & 64.7 & 68.6 & 6.9 & 9.8 & 11.8 & 13.7 \\
\bottomrule
\end{tabular}
\end{table}

Surprisingly, smaller models like GPT-4o mini and Claude 3.5 Haiku out-of-the-box passed more tests than their more powerful counterparts (Claude Sonnet and GPT-4o). One of the observed reasons is that smaller models tend to write scalar C++ code by default (i.e., regular C++ with for loops instead of vector intrinsics). Stronger models may try to write vectorized code at first, but end up hallucinating functions as shown in Fig. \ref{fig:complexabs}. With compiler feedback these models will often fall back to scalar solutions, which will allow them to pass more tests.

DeepSeek V3, Claude 3.7 Sonnet and GPT-4o benefitted greatly from successive compilation attempts, whereas others saw marginal benefits. DeepSeek V3 in particular was trying to include ``adf.h'' in nearly every first attempt, which does not exist in the open source tool suite used within the evaluation harness, which is why after a single recompilation its pass rate shot up to 42\%.

\subsection{Performance}

\begin{figure}[!h]
    \centering
    \includegraphics[width=\columnwidth]{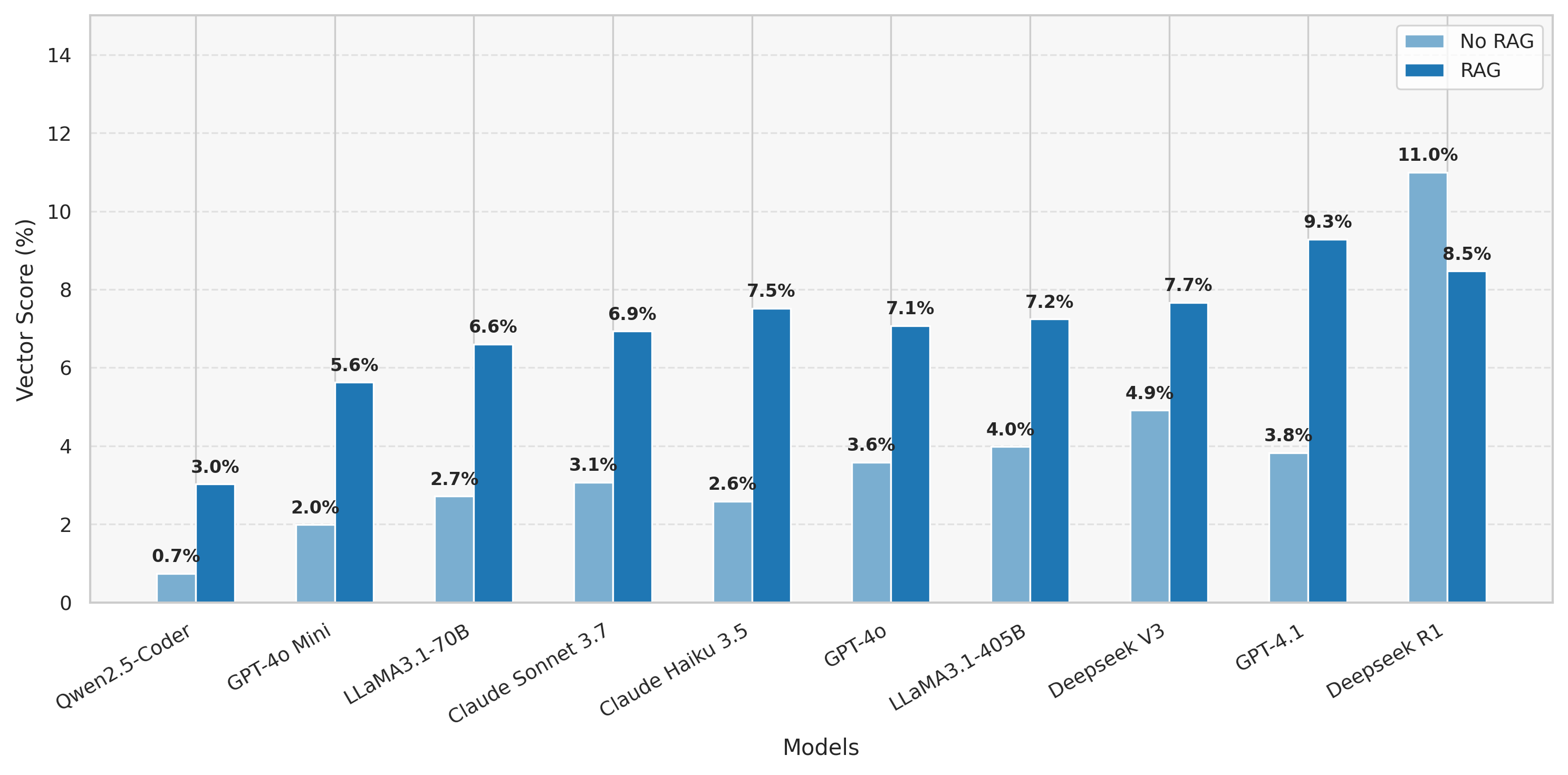}
    \caption{Vectorization results}
    \label{fig:vector_results}
\end{figure}

The efficiency of generated kernels is calculated by dividing the number of cycles that are used for VPU execution by the total number of cycles it takes to process the test input on the NPU. While not perfect, this is a good proxy for how well the kernel is utilizing the VPU. Kernels that do not pass functional tests are evaluated as 0\%. The average score for all kernels per model is displayed in Fig. \ref{fig:vector_results}. While the scores seem very low it should be noted that SoTA open source kernels\cite{mlir_aie} will typically see a vectorization factor of 10-30\% as shown in Appendix \ref{appendix_a}.

To guide model responses towards vectorized code we introduced vectorized kernel RAG into the code generation pipeline. Introducing RAG composed of openly available AIE kernels into the prompts improved VPU utilization across the board, with GPT-4.1 seeing the sharpest improvement in vectorization score. Curiously the only model which saw a decrease in average VPU utilization was DeepSeek R1, which still holds the best score without using vectorized examples. The reason it performed worse is because some kernels in our database had compiler-specific pragmas which are different to the pragmas one would use with LLVM-AIE. Refer to Appendix \ref{appendix_d} for solution examples and how they differ.

With compiler feedback, DeepSeek V3 and GPT-4.1 were among the best models at generating vectorized code, even though they scored quite low on correctness testing. This follows the trend of powerful models being more ``inventive'' when trying to write performant code, however due to lack of deep knowledge of NPU programming environments were more prone to mistakes.

\subsection{LLM Failure Analysis}

\begin{figure*}[!h]
\centering
\begin{minipage}{0.48\textwidth}
    \centering
    \includegraphics[width=\linewidth]{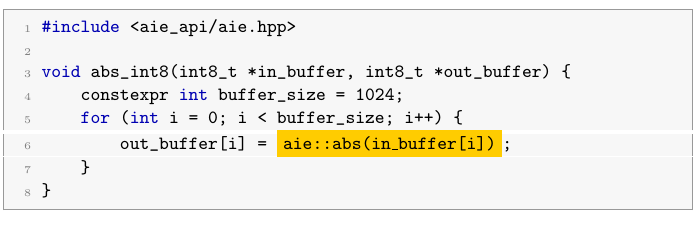}
    \subcaption{Looping one element at a time (GPT-4o)}
    \label{fig:abs_int8}
    \vspace{10pt}

    \includegraphics[width=\linewidth]{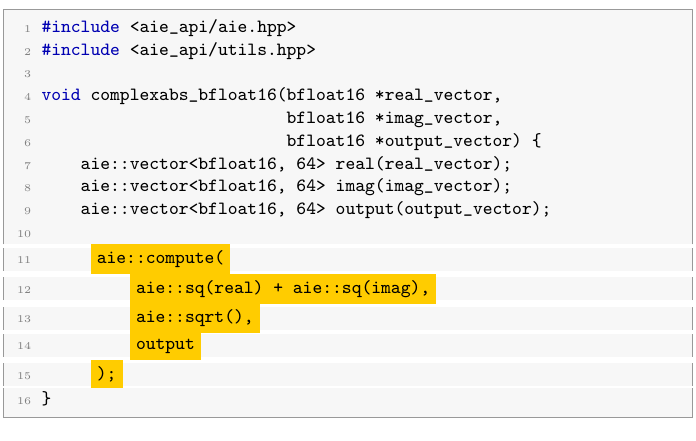}
    \subcaption{Hallucinated non-existent API (Llama-3.1 70b)}
    \label{fig:complexabs}
    
\end{minipage}
\hfill
\begin{minipage}{0.48\textwidth}
    \centering
    \vspace{-46pt}
    \includegraphics[width=\linewidth]{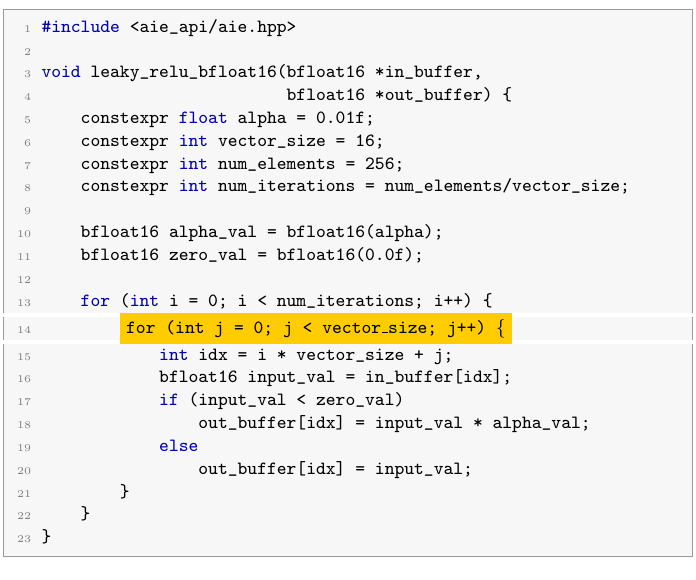}
    \subcaption{Aware of vectors and how to chop up the data but processing in scalar loops (Claude 3.7 Sonnet)}
    \label{fig:leaky_relu}
\end{minipage}
\caption{Examples of LLM errors in vectorized kernel generation: scalar loop with conditional (claude-3-7-sonnet), incomplete vectorization (gpt-4o), and hallucinated API usage (llama-3.1-70b).}
\label{fig:llm-failure-modes}
\end{figure*}

The LLMs that passed many functional correctness tests ended up writing very ineffient solutions as shown in Fig \ref{fig:abs_int8} -- these kernels do not utilize the NPU hardware to the fullest. Incorrect solutions will fail compilation often due to hallucinations as shown in Fig \ref{fig:complexabs} or misunderstanding of how to use vector APIs as shown in Fig \ref{fig:leaky_relu} where the model has successfully used the AIE APIs but was doing it in a loop one element at a time.

There seems to be baked in knowledge in these models however, Claude 3.7 Sonnet (Fig \ref{fig:leaky_relu}) does have a notion of vector\_size and that it should be chunking the input buffer into vectors of 16 elements. With more advanced prompting and more quality examples these models have the potential to write efficient NPU kernels.

\section{Limitations}

In this study we have primarily targeted the AIE architecture found in AMD NPUs. While methodology is designed to be extensible to other accelerator families and even within the same AIE family as new devices are released updated with latest intrinsics and API improvements.

The presented results have been achieved using one compiler backend, however there are multiple options for just the AIE. And as we saw from the DeepSeek R1 result, certain optimizations in the code are compiler specific which will be an interesting challenge when generalizing across NPU architectures and programming platforms.

Since NPUs are a relatively new class of device and the programming model is not as established other accelerators there is still a lack of quality open source kernels to establish a human baseline. We include a couple examples in Appendix \ref{appendix_a} to contrast our results, and note that current SoTA is not higher than 30\%, though this will be highly kernel dependent.

Our evaluation covered a range of popular LLMs, but more models could be tested. All results are based on greedy decoding with a single pass; while pass@k evaluations may yield higher scores, one-shot decoding still offers meaningful insight into current LLM capabilities for NPU kernel generation.

\section{Future work}

NPUs are still a new class of accelerator and its impressive how good some models already are at writing code for them. This is only the beginning and it will be exciting to see how new techniques like code generating agents will improve upon this benchmark.

While the work presented in this paper focused on one NPU architecture, this methodology could easily extend across different vendors and families of NPUs. The Python behavioral models can be re-used along with the PromptConstructor class to generate datasets targeting any programming model. We plan to use the same methodology to extend this work to other accelerator families.

\section{Conclusions}

We have presented NPUEval, the first benchmark to systematically evaluate LLMs in their ability to generate NPU kernel code. NPUEval includes a comprehensive evaluation harness with cycle-accurate performance metrics and a reference code generation pipeline.

Our results show that most LLMs can readily generate scalar code, but struggle to produce optimized, vectorized solutions. Interestingly, smaller models (e.g., GPT-4o mini, Claude 3.5 Haiku) tend to favor functional but unoptimized code, while stronger models (e.g., GPT-4o, Claude 3.7 Sonnet, DeepSeek R1) attempt to optimize and risk hallucinations, leading to lower functional scores.

We believe NPUEval provides a valuable foundation for advancing LLM-driven accelerator kernel generation. We hope this benchmark will become the standard for measuring and improving LLM-based code generation for emerging hardware architectures.

\section*{Acknowledgments}

We would like to thank our colleagues Mario Ruiz Noguera, Thomas Papatheodore, Stephen Neuendorffer, Jack Lo, and Joseph Melber from AMD for their expertise and technical advice. Many thanks to Patrick Lysaght for supporting the early development stages of this project.

Additionally we want to thank Lakshya A Agrawal and Matei Zaharia (UC Berkeley) for their feedback in the development of the benchmark.

\begingroup
\let\itshape\upshape
\bibliographystyle{unsrt}
\bibliography{refs}
\endgroup


\newpage
\appendix

\section{Examples of vectorized kernels} \label{appendix_a}

\definecolor{verylightgray}{rgb}{0.97, 0.97, 0.97}
\definecolor{commentcolor}{rgb}{0.0, 0.5, 0.0}
\definecolor{keywordcolor}{rgb}{0.0, 0.0, 0.7}
\definecolor{stringcolor}{rgb}{0.7, 0.0, 0.0}
\definecolor{framecolor}{rgb}{0.6, 0.6, 0.6}

\lstset{
  language=C++,
  basicstyle=\small\ttfamily,
  backgroundcolor=\color{verylightgray}, 
  breaklines=true,
  frame=single,
  rulecolor=\color{framecolor},
  numbers=left,
  numberstyle=\tiny\color{gray},
  numbersep=5pt,
  xleftmargin=15pt,
  framexleftmargin=15pt,
  commentstyle=\color{commentcolor},
  keywordstyle=\color{keywordcolor},
  stringstyle=\color{stringcolor},
  captionpos=b,
  stepnumber=1,
  showstringspaces=false,
  linewidth=\linewidth,
  columns=flexible,
  keepspaces=true,
  basewidth=0.5em
}

Here we provide example vectorized kernels available in open source repositories like \url{https://github.com/xilinx/mlir-aie}. While the same functionality can be achieved writing regular C++ loops, vectorized code is more specialized and uses intrinsics and AIE APIs.

Listing \ref{lst:vectoradd_example} shows an open source implementation of an elementwise add kernel, which uses Chess compiler pragmas and AIE APIs to perform vector operations.

\begin{lstlisting}[label=lst:vectoradd_example, caption=Elementwise Add example (Vector score: 13\%)]
#define NOCPP

#include <stdint.h>
#include <stdio.h>
#include <stdlib.h>
#include <type_traits>

#include <aie_api/aie.hpp>

template <typename T_in, typename T_out, const int N>
void eltwise_add(T_in *a, T_in *b, T_out *c) {
  for (int i = 0; i < N; i++) {
    c[i] = a[i] + b[i];
  }
}

template <typename T_in, typename T_out, const int N>
void eltwise_vadd(T_in *a, T_in *b, T_out *c) {

  constexpr int vec_factor = 16;
  event0();
  T_in *__restrict pA1 = a;
  T_in *__restrict pB1 = b;
  T_out *__restrict pC1 = c;
  const int F = N / vec_factor;
  for (int i = 0; i < F; i++)
    chess_prepare_for_pipelining chess_loop_range(16, ) {
      aie::vector<T_in, vec_factor> A0 = aie::load_v<vec_factor>(pA1);
      pA1 += vec_factor;
      aie::vector<T_in, vec_factor> B0 = aie::load_v<vec_factor>(pB1);
      pB1 += vec_factor;
      aie::vector<T_out, vec_factor> cout = aie::add(A0, B0);
      aie::store_v(pC1, cout);
      pC1 += vec_factor;
    }
  event1();
}

extern "C" {

void eltwise_add_bf16_scalar(bfloat16 *a_in, bfloat16 *b_in, bfloat16 *c_out) {
  eltwise_add<bfloat16, bfloat16, 1024>(a_in, b_in, c_out);
}

void eltwise_add_bf16_vector(bfloat16 *a_in, bfloat16 *b_in, bfloat16 *c_out) {
  eltwise_vadd<bfloat16, bfloat16, 1024>(a_in, b_in, c_out);
}

} // extern "C"
\end{lstlisting}

In listing \ref{lst:vectorizedconv_example} is an optimized version of a conv2d kernel, which achieves 30\% vectorization score.

\begin{lstlisting}[label=lst:vectorizedconv_example, caption=Conv2D example (Vector score: 30\%)]
#define NOCPP

#include <stdint.h>
#include <stdio.h>
#include <stdlib.h>

#include <aie_api/aie.hpp>

#define REL_WRITE 0
#define REL_READ 1

#ifdef SCALAR

const int32_t SMAX = 127;
const int32_t SMIN = 128;

#ifdef INT8_ACT
//*****************************************************************************
// conv2d 1x1 - scalar
// act: int8, wts: int8, out: int8
//*****************************************************************************
void conv2dk1_i8_scalar(int8_t *input, int8_t *kernels, int8_t *output,
                        const int32_t input_width, const int32_t input_channels,
                        const int32_t output_channels, const int scale) {
  event0();

  int x, ic, oc, ic8, oc8;
  // scale=-17;
  for (oc = 0; oc < output_channels / 8; oc++) {
    for (x = 0; x < input_width; x++) { // col of output image
      for (oc8 = 0; oc8 < 8; oc8++) {
        int sum = 0;
        int sum_srs = 0;

        for (ic = 0; ic < input_channels / 8; ic++) {
          for (ic8 = 0; ic8 < 8; ic8++) {
            int val = input[(ic * input_width * 8) + (x * 8) + ic8];
            int k = kernels[(oc * (input_channels / 8) * 64) + (ic * 64) +
                            (ic8 * 8) + oc8];
            sum += val * k;
          }
        }

        // sum_srs=sum>>scale;
        sum_srs = (sum + (1 << (scale - 1))) >> scale;
        sum_srs = (sum_srs > SMAX) ? SMAX : (sum_srs < -SMIN) ? -SMIN : sum_srs;
        // sum_srs = input[(oc*input_width*8) + (x*8) + oc8];
        output[(oc * input_width * 8) + (x * 8) + oc8] = sum_srs;
      }
    }
  }

  event1();
}
#endif // INT8_ACT

#else // Vector

#ifdef INT8_ACT

//*****************************************************************************
// conv2d 1x1 - vector
// act: int8, wts: int8, out: uint8
//
// Assume IC >= 16 as that gives ideal inner loop schedule
//
// TODO - Restricting input_width is mutiple of 32
// Because each VMAC works on 4 inputs at a time and we store intermediate
// results in 8 accumulators, having input_width be a multiple of 4*8=32 is
// ideal. However, we should be able to support input_width that is only a
// multiple of 4 but there is some strange scheduling happening now so for
// now, we do not.
//*****************************************************************************
void conv2dk1_i8_vector(int8_t *input, int8_t *kernels, int8_t *output,
                        const int32_t input_width, const int32_t input_channels,
                        const int32_t output_channels, const int scale) {
  event0();

  using MMUL4x8x8 = aie::mmul<4, 8, 8, int8, int8>;
  ::aie::set_saturation(
      aie::saturation_mode::saturate); // Needed to saturate properly to uint8
  ::aie::set_rounding(aie::rounding_mode::symmetric_inf); // Needed to saturate
                                                          // properly to uint8

  int8_t *restrict out_ptr = output;

  const int scaleT = scale;

  MMUL4x8x8 acc_tmp[8];
  for (int x = 0; x < 8; x++) {
    acc_tmp[x] = aie::zeros<acc32, 32>();
  }

  // TODO Keeping this variable gives a wrong behavior and bad schedule!
  const int iw = input_width;
  const int iw_32 = (input_width / 4) / 8;

  // const int iw_32_rem = (input_width / 4) % 8;
  // const int iw_32_rem = (32 / 4) % 8;
  assert((input_width / 4) % 8 == 0);
  const int iw_32_rem = 0; // TODO - See restriction

  assert((input_channels / 8) > 2); // Assume IC >= 16

  if (iw_32 > 0) {

    for (int oc = 0; oc < (output_channels / 8); oc++) {
      for (int iw_32c = 0; iw_32c < iw_32; iw_32c++) {
        for (int ic = 0; ic < (input_channels / 8); ic++)
          chess_prepare_for_pipelining chess_loop_range(2, ) {
            aie::vector<int8, 64> in_b = aie::load_v<64>(kernels);
            kernels += 64; // wts ic0..7(oc0..7)

            for (int x = 0; x < 8; x++) {
              aie::vector<int8, 32> in_a = aie::load_v<32>(input);
              input += 32; // act oc0..3(ic0..7)
              acc_tmp[x].mac(in_a, in_b);
            }
            input += (iw * 8) - 256; // Move to next ic/8 position
          }
        // input ptr just moves to next section
        for (int xx = 0; xx < 8; xx++) {
          aie::vector<int8, 32> o1 = acc_tmp[xx].to_vector<int8>(scaleT);
          aie::store_v(out_ptr, o1);
          out_ptr += 32;
          acc_tmp[xx] = aie::zeros<acc32, 32>();
        }
        input -= ((input_channels / 8) * iw * 8) -
                 256; // reset to next input_width/32 block
        kernels -=
            (input_channels / 8) * 64; // reset kernel back to beginning of ic/8
      }
      input -= (iw_32) * 256; // 8*32, reset beginning of input ptr
      kernels += (input_channels / 8) * 64; // move to next oc/8 weights
      out_ptr += (iw_32_rem *
                  32); // move to next oc/8 (skip remainder section if present)
    }

  } // if(iw_32 > 0) {

  if (iw_32_rem > 0) {

    const int ocs = output_channels;
    const int ics = input_channels;

    for (int oc = 0; oc < (ocs / 8); oc++) {
      for (int ic = 0; ic < (ics / 8); ic++)
        chess_prepare_for_pipelining chess_loop_range(2, ) {
          aie::vector<int8, 64> in_b = aie::load_v<64>(kernels);
          kernels += 64; // wts ic0..7(oc0..7)

          for (int x = 0; x < iw_32_rem; x++) {
            aie::vector<int8, 32> in_a = aie::load_v<32>(input);
            input += 32; // act oc0..3(ic0..7)
            acc_tmp[x].mac(in_a, in_b);
          }
          input += (iw * 8) - (iw_32_rem * 32); // Move to next ic/8 position
        }
      // input ptr just moves to next section
      for (int xx = 0; xx < iw_32_rem; xx++) {
        aie::vector<int8, 32> o1 = acc_tmp[xx].to_vector<int8>(scaleT);
        aie::store_v(out_ptr, o1);
        out_ptr += 32;
        acc_tmp[xx] = aie::zeros<acc32, 32>();
      }
      // input   -= ((ics-1)/8)*(iw*8)+(iw_32_rem*32); // reset to beginning of
      // input ptr for remainder
      input -= 448; // reset to beginning of input ptr for remainder
      // kernel ptr already at next oc/8
      out_ptr += (iw * 8) -
                 (iw_32_rem *
                  32); // move to next oc/8 (skip remainder section if present)
    }

  } // if(iw_32_rem > 0)

  event1();
}
#endif // INT8_ACT
#endif // Vector

//*****************************************************************************
// conv2d 1x1 wrappers
//*****************************************************************************
extern "C" {

#ifdef SCALAR

#ifdef INT8_ACT

void conv2dk1_i8(int8_t *input, int8_t *kernels, int8_t *output,
                 const int32_t input_width, const int32_t input_channels,
                 const int32_t output_channels, const int scale) {
  conv2dk1_i8_scalar(input, kernels, output, input_width, input_channels,
                     output_channels, scale);
}
#endif // INT8_ACT
#else  // Vector

#ifdef INT8_ACT

void conv2dk1_i8(int8_t *input, int8_t *kernels, int8_t *output,
                 const int32_t input_width, const int32_t input_channels,
                 const int32_t output_channels, const int scale) {
  conv2dk1_i8_vector(input, kernels, output, input_width, input_channels,
                     output_channels, scale);
}
#endif // INT8_ACT
#endif // Vector
} // extern "C"
\end{lstlisting}

\section{Prompt construction ablation study} \label{appendix_b}

\begin{figure}[!h]
    \begin{subfigure}[b]{.5\textwidth}
        \centering
        \includegraphics[width=\textwidth]{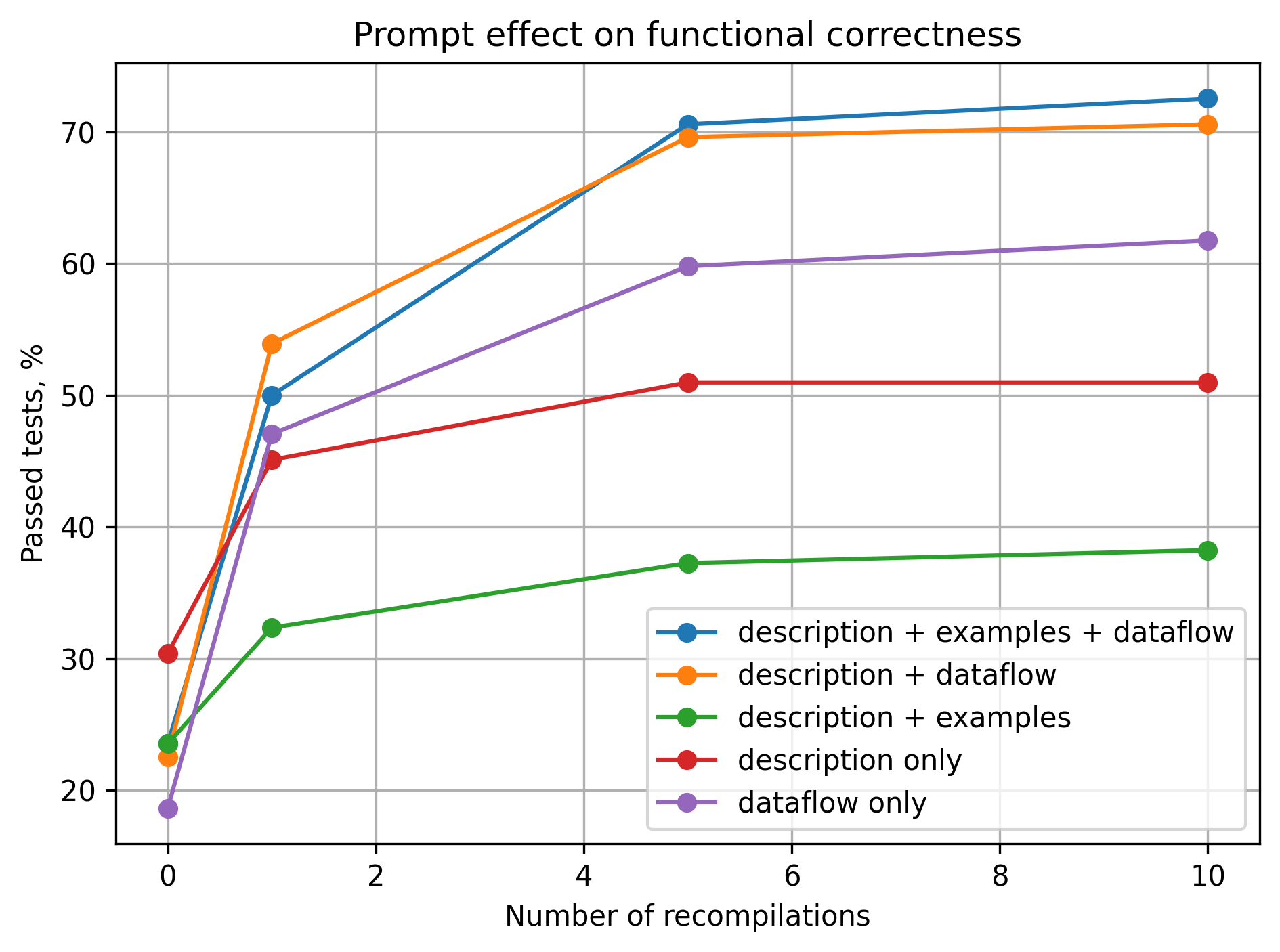}
        \caption{Functional correctness results}
    \end{subfigure}
    \begin{subfigure}[b]{.5\textwidth}
        \centering
        \includegraphics[width=\textwidth]{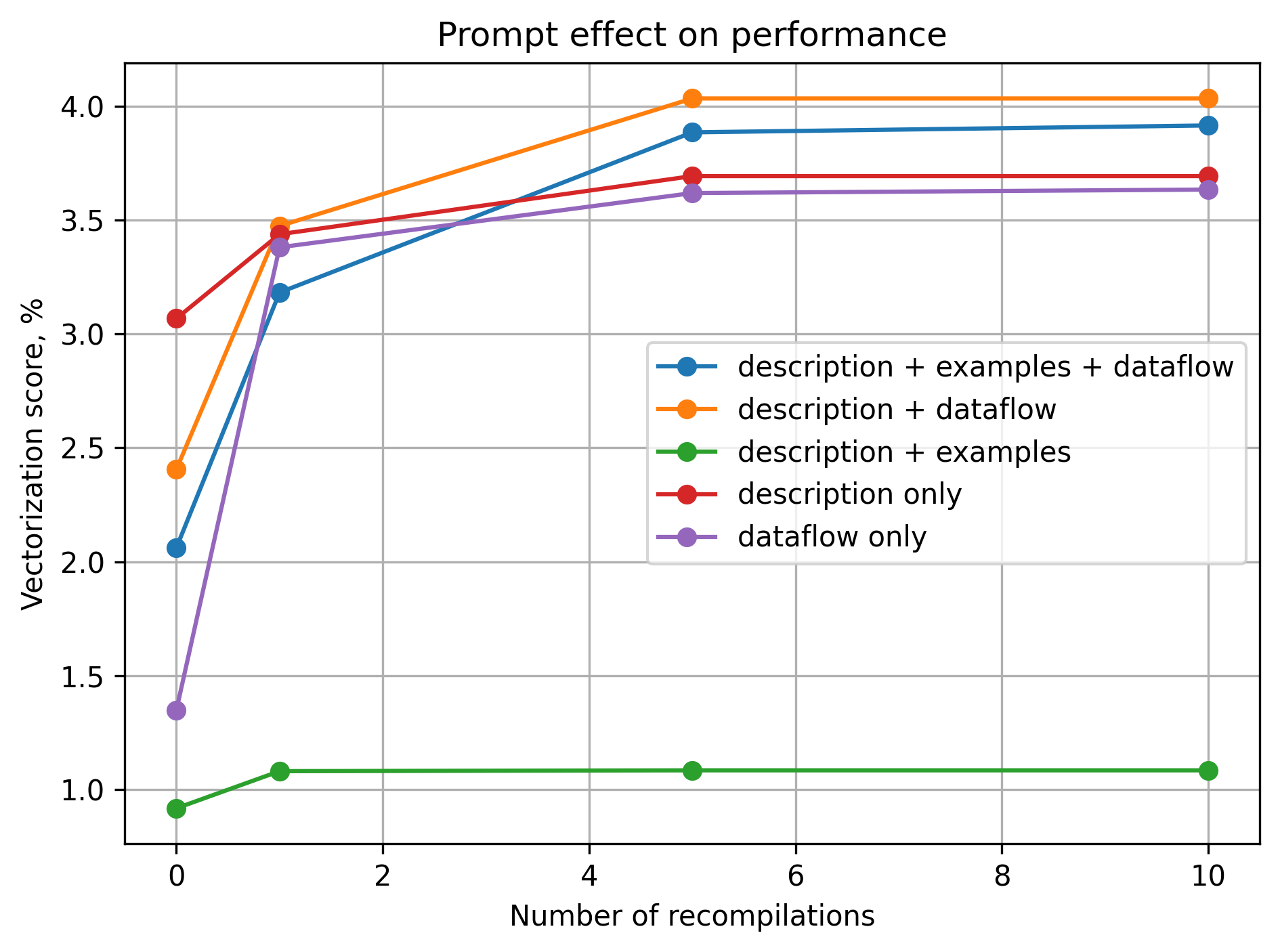}
        \caption{Vectorization results}
    \end{subfigure}
    \caption{Simple example of vectorizing a passthrough kernel.}
    \label{fig:dataset_ablation}
\end{figure}

Here we evaluate a number of prompt configurations when determining the final structure for NPUEval prompts. Figure \ref{fig:dataset_ablation} illustrates how well GPT-4.1 scores with different parts of the docstring being omitted.

\begin{table}[!h]
  \caption{Prompt construction study}
  \label{sample-table}
  \centering
  \begin{tabular}{lll}
    \toprule
    Name     & Description      \\
    \midrule
    Description + examples + dataflow & Includes everything   \\
    Description + dataflow            & Omits inline examples \\
    Description + examples            & Omits input/output size information   \\
    Description only                  & Omits inline examples and input/output/sizes \\
    Dataflow only                     & Omits description and inline examples \\
    \bottomrule
  \end{tabular}
\end{table}

Unsurprisingly, omitting dataflow information had the most grevious effect on the overall perforamnce of the LLM. However, additionally removing the inline examples and only keeping the description significantly increased the number of passing functional tests. In this case without having ground truth expected input shapes, the model relies on the example shapes which are not representative of the actual test sizes (since it would be impractical to bake in large arrays of numbers into the prompt), this causes it to use the example shapes or hallucinate expected parameters to design for.

If expected input buffer and parameter information, along with a high level description of the kernel are present then adding additional inline examples of algorithmic behavior is actually beneficial in passing more tests. While the vectorization score is slightly higher without the examples, the delta is less than 0.1 which makes it close to negligible.

If you choose to exclude examples or modify the prompts in any way this is fully supported by the codebase.

\newpage

\section{Performance vs number of iterations} \label{appendix_c}

We test a number of LLMs for up to 10 compilation attempts and see a trend of diminishing returns at around 5 turns. Models like DeepSeek V3, Claude Sonnet and GPT-4.1 are especially receptive to the compiler feedback and see a sharp increase in passing tests.

\begin{figure}[!h]
    \centering
    \begin{subfigure}[b]{.75\textwidth}
        \centering
        \includegraphics[width=\textwidth]{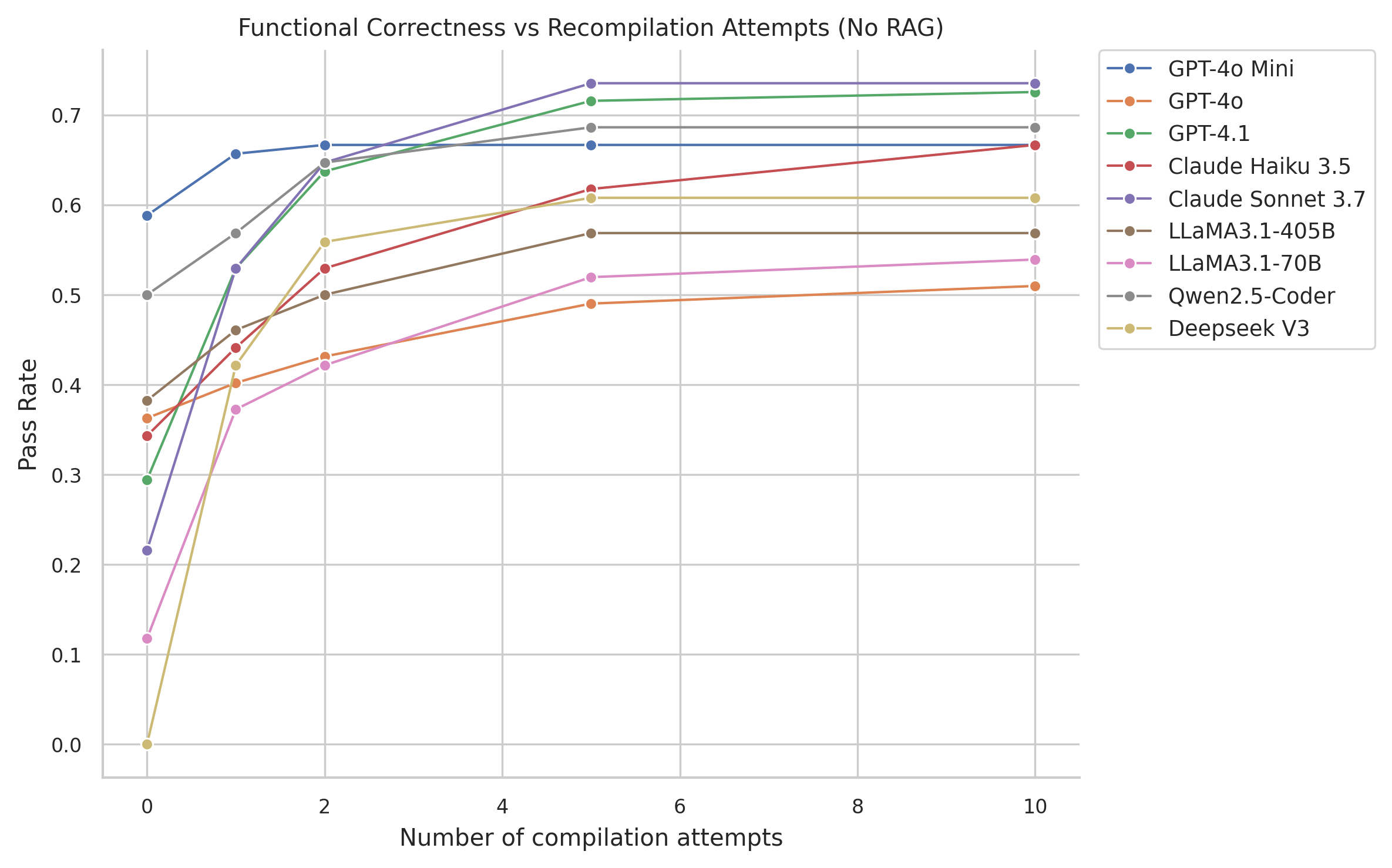}
        \caption{Test pass rate as number of recompilations increases}
        \label{fig:10_recompiles}
    \end{subfigure}
    \vfill
    \begin{subfigure}[b]{.75\textwidth}
        \centering
        \includegraphics[width=\textwidth]{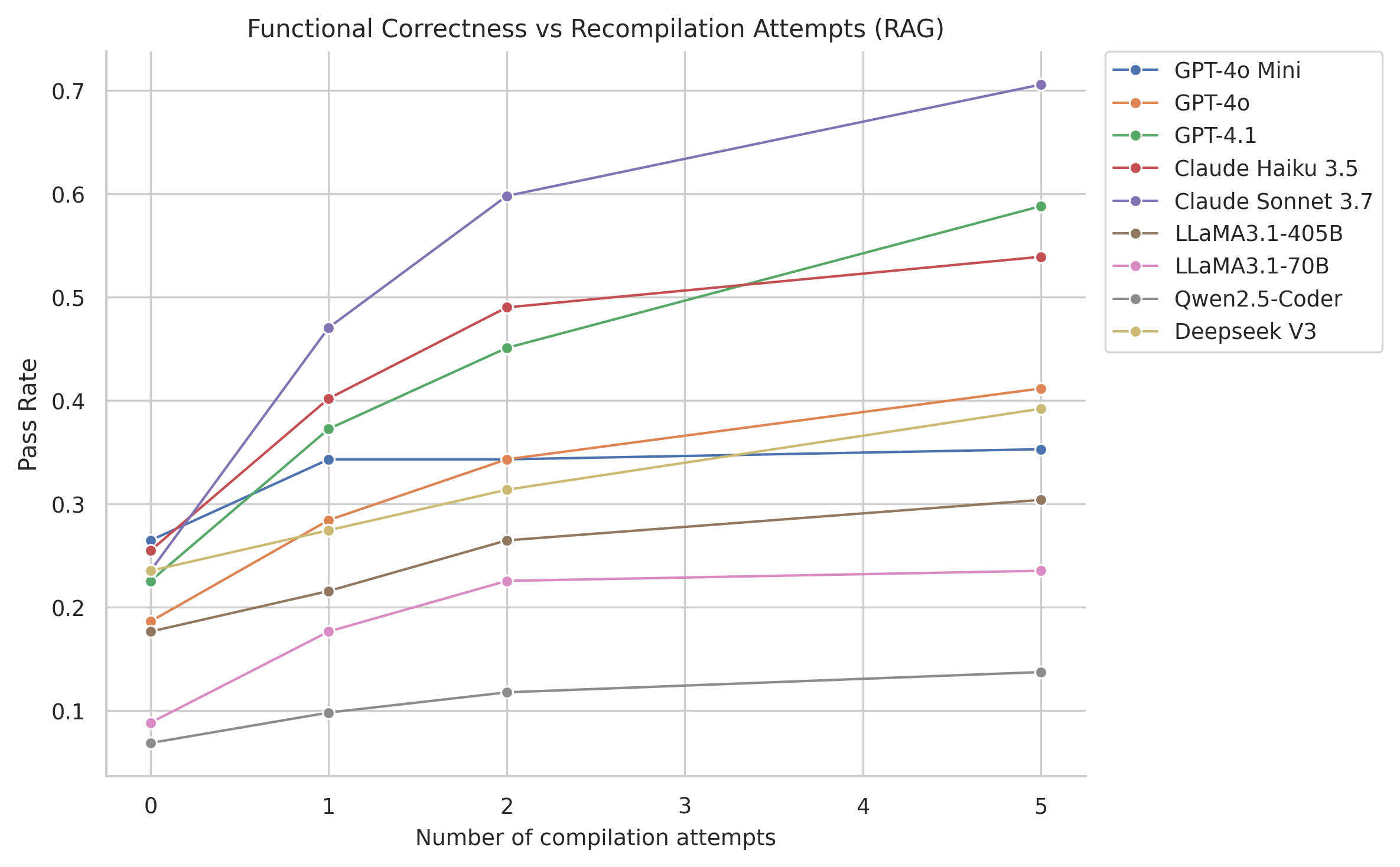}
        \caption{Visualization of pass rate with RAG included}
        \label{fig:10_recompiles_rag}
    \end{subfigure}
    \caption{Number of compiler iterations vs pass rate.}
    \label{fig:appendix_recompilations}
\end{figure}

\section{Closer Look at DeepSeek R1 Results} \label{appendix_d}

A surprising result from our study was the decrease in VPU utilization from the DeepSeek R1 model when introduced to vectorized code samples. This was the only model that decreased in performance out of the tested LLMs. 

Looking at the solutions it produced without any RAG in Listing \ref{lst:deepseek_norag} and comparing with Listing \ref{lst:deepseek_rag} with RAG we observe that it is using compiler-specific pragmas. It seems like the R1 model is well versed in how to use pragmas to do low-level optimization like loop unrolling and the examples in our RAG database had code specifically tailored to the Chess compiler, which uses a different set of pragmas to LLVM-AIE (Peano). 

\definecolor{verylightgray}{rgb}{0.97, 0.97, 0.97}
\definecolor{commentcolor}{rgb}{0.0, 0.5, 0.0}
\definecolor{keywordcolor}{rgb}{0.0, 0.0, 0.7}
\definecolor{stringcolor}{rgb}{0.7, 0.0, 0.0}
\definecolor{framecolor}{rgb}{0.6, 0.6, 0.6}

\lstset{
  language=C++,
  basicstyle=\small\ttfamily,
  backgroundcolor=\color{verylightgray}, 
  breaklines=true,
  frame=single,
  rulecolor=\color{framecolor},
  numbers=left,
  numberstyle=\tiny\color{gray},
  numbersep=5pt,
  xleftmargin=15pt,
  framexleftmargin=15pt,
  commentstyle=\color{commentcolor},
  keywordstyle=\color{keywordcolor},
  stringstyle=\color{stringcolor},
  captionpos=b,
  stepnumber=1,
  showstringspaces=false,
  linewidth=\linewidth,
  columns=flexible,
  keepspaces=true,
  basewidth=0.5em
}

\begin{minipage}{4.5in}
\begin{lstlisting}[label=lst:deepseek_norag, caption=Solution without RAG (Vector score: 47\%)]
#include <aie_api/aie.hpp>

void relu_bfloat16(bfloat16 *in_buffer, bfloat16 *out_buffer) {
    constexpr int VECTOR_SIZE = 16;
    constexpr int NUM_VECTORS = 256 / VECTOR_SIZE;
    auto zero_vec = aie::broadcast<bfloat16, VECTOR_SIZE>(bfloat16(0.0f));

    for (int i = 0; i < NUM_VECTORS; ++i) {
        aie::vector<bfloat16, VECTOR_SIZE> vec = aie::load_v<VECTOR_SIZE>(in_buffer);
        vec = aie::max(vec, zero_vec);
        aie::store_v(out_buffer, vec);
        in_buffer += VECTOR_SIZE;
        out_buffer += VECTOR_SIZE;
    }
}
\end{lstlisting}
\end{minipage}

\begin{minipage}{4.5in}
\begin{lstlisting}[label=lst:deepseek_rag, caption=Solution with RAG (Vector score: 26\%)]
#include <aie_api/aie.hpp>

void relu_bfloat16(bfloat16 *in_buffer, bfloat16 *out_buffer) {
    const int v_factor = 32;
    v32bfloat16 zeroes = broadcast_zero_to_v32bfloat16();

    for (int i = 0; i < 256; i += v_factor) {
        chess_prepare_for_pipelining
        chess_loop_range(8, 8) {
            v32bfloat16 input = *(v32bfloat16 *)(in_buffer + i);
            v32bfloat16 output = max(input, zeroes);
            *(v32bfloat16 *)(out_buffer + i) = output;
        }
    }
}
\end{lstlisting}
\end{minipage}

While Peano will not throw an errow after encountering Chess-specific pragmas it will ignore them during optimization passes. Without being shown these examples the LLM (Listing \ref{lst:deepseek_norag}) will fall back to standard AIE APIs which are already abstracting optimized operations without worry of lower level intrinsics or pragmas.

The way to address this in future work would be to have compiler backend-specific RAG. This information could also be potentially conveyed via sytem prompts and will likely become a necessity as more NPU architectures get released, each with its own unique set of intrinsics and compiler-specific optimization paths.



\end{document}